\documentclass{cernart}
\usepackage{times}
\usepackage{float}
\usepackage{amssymb}
\usepackage{epsfig}

\def\numtonut{{\mathrm\nu}_{\mathrm\mu}\rightarrow{\mathrm\nu}_{\mathrm\tau}}

\def\sin2t{sin^22\theta_{\mu\tau}}

\def\dm2{\Delta\mathrm{m}^2_{\mathrm\mu \tau}}

\begin{document}
\begin{titlepage}
\title{\large New results on the $\numtonut$ oscillation search with the CHORUS
  detector}
\begin{Authlist}
{\large \bf The CHORUS Collaboration}

\end{Authlist}
\vskip 2cm

\begin{abstract}
  The present results on the $\numtonut$ oscillation search by the CHORUS
  experiment at CERN are summarised. A fraction of the neutrino interactions
  collected in 1994-1995-1996 by the CHORUS experiment has been analysed,
  searching for $\nu_{\tau}$ charged current interactions followed by the
  $\tau$ lepton decay into a negative hadron or into a muon. A sample of 68,156
  events with an identified final state muon and 7,206 events without an
  identified muon in the final state have been located in the emulsion target.
  Within the applied cuts, no $\nu_{\tau}$ candidate has been found.  This
  result leads to a 90\% C.L. limit $P(\numtonut)< 6.0\cdot 10^{-4}$ on the
  mixing probability.
\end{abstract}
\submitted{Contributed paper at the XXIX International Conference on High
  Energy Physics\\
23-28 July, 1998, Vancouver, BC, Canada.
}
\end{titlepage}
\twocolumn
\setlength{\textheight}{240mm}

The CHORUS experiment has recently reported~\cite{1mupap,0mupap,nu98} a limit
on $\nu_{\mu}-\nu_{\tau}$ oscillation obtained by the analysis of a subsample
of neutrino interactions, taken in 1994-1995, both with and without an
identified ${\mu}^-$ in the final state. This paper contains an update of the
statistics of the subsample of events with an identified $\mu^-$ in the final
state.

The experimental setup and the characteristics of the CERN wide band neutrino
beam are summarised in~\cite{1mupap} and described in more details in
~\cite{detpap}.

\section{The apparatus}

The $"hybrid"$ CHORUS apparatus combines a 770 kg nuclear emulsion target with
various electronic detectors : a scintillating fibre tracker system, trigger
hodoscopes, an air-core magnet, a lead/scintillator calorimeter and a muon
spectrometer. 

Details about the experimental setup and the performances of the sub-detectors
can be found in~\cite{detpap}.

\section{ Data collection and event selection  }

In the 1994-1997 period, CHORUS has collected 2,271,000 triggers corresponding
to $5.06 \ 10^{19}$ protons on target. Of these, 458,601 have a muon identified
in the final state (the so called $1\mu$ events) and 116,049 do not (the so
called $0\mu$ events) and a vertex position compatible with one of the four
emulsion target stacks.

All tracks, associated to the interaction vertex, with an angle less than 0.4
rad with the beam axis and, in view of the large emulsion background of muons
originating from a nearby secondary target, bigger than 0.05 rad from the
direction of this target are extrapolated downstream and selected for further
analysis. These tracks are searched for in the emulsion if their charge is
negative and their momentum is in
the range $0\leq p_h \leq 20$~GeV/c and $0\leq p_\mu \leq 30$~GeV/c for
hadrons and muons, respectively.
    
It should be noted that use has not been made of the energy deposition in the
calorimeter to reject electrons or unidentified muons. The contribution of the
$\tau^-$ leptonic decay modes to the $0\mu$ data sample is taken into account
in the evaluation of the sensitivity.
 
\begin{table*}
\caption{\em {Current status of the CHORUS analysis}}
\label{tab3}
 \begin{tabular}{llll}
\\
\hline
\hline
   & 1994 & 1995 & 1996 \\ 
\hline
Protons on target & $0.81\cdot10^{19}$ & $1.20\cdot10^{19}$  &  $1.38\cdot10^{19}$ \\
 \hline
Emulsion triggers & 422,000 & 547,000 & 617,000\\
 \hline
$1\mu$ to be scanned & 66,911 & 110,916 & 129,669 \\
\hline
$0\mu$ to be scanned & 17,731 & 27,841 & 32,548 \\
\hline
$1\mu$ scanned so far & 42,154 & 49,912 & 72,615 \\
\hline
$0\mu$ scanned so far & 8,908 & 12,635 & - \\
\hline
$1\mu$ vertex located (in the 33 most upstream plates) &18,286  &20,642 & 30,128 \\
\hline
$0\mu$ vertex located (in the 33 most upstream plates) &3,401  &3,805 & 0   \\
 \hline
  \end{tabular}
\end{table*} 

\section{Scanning procedure  }
\subsection{Vertex location }

The various steps leading to the plate containing the vertex by means of fully
automatised microscopes are identical to those described
in~\cite{1mupap,0mupap}. They are independently applied to all the selected
tracks in the event, the muon for the $1\mu$ events, all the negative tracks
for the $0\mu$ events.  A track which has been found in the interface emulsion
sheets is followed upstream in the target emulsion, using track segments
reconstructed in the most upstream 100~$\mu$m of each plate, until the track
disappears. This plate is referred to as the vertex plate, since it
should contain the primary neutrino vertex or the secondary (decay) vertex from
which the track originates.  The three most downstream plates of each stack are
used to validate the matching with the interface emulsion sheets and are not
considered as possible vertex plates. The scanning results are summarised in
Table~\ref{tab3}.

The mean efficiency of this scan-back procedure is found to be $\sim 32\%$ and
$\sim 42\%$ for $0\mu$ and $1\mu$ events, respectively.  A detailed simulation
of the scanning criteria shows that the difference mainly reflects the poor\-er
quality of the hadron track predictions at the interface emulsion sheets,
because of the difficulty of reconstruction inside a dense ha\-dro\-nic
sho\-wer and the larger multiple scattering owing to the lower average
momentum.
  
\subsection{ Decay search }

Once the vertex plate is defined, automatic microscope measurements are
performed to select the events potentially containing a decay topology (kink).
Different algorithms have been applied, as a result of the pro\-gress in the
scanning procedures and of the improving performance in speed of the scanning
devices. They are described in~\cite{1mupap,0mupap} and briefly recalled here.
  
In the first procedure the event is selected either when the scan-back track
has a significant impact parameter with respect to the other predicted tracks
or when the change in the scan-back track direction between the vertex plate
and the exit from the emulsion corresponds to an apparent transverse momentum,
$p_{T}$, larger than 250 MeV/c.  For the selected e\-ven\-ts and for those
with only one predicted track, digital images of the vertex plate are recorded
and are analysed off-line for the presence of a kink.
  
The second procedure is restricted to the search of {\em long} decay paths. In
that case the vertex plate is assumed to contain the decay vertex of a charged
parent particle produced in a more upstream plate. With this procedure only
kink angles larger than 0.025 rad are detected.
  
For the events selected by either one of these procedures, a computer assisted
eye-scan is performed to assess the presence of a secondary vertex and measure
accurately its topology. A $\tau^-$ decay candidate must satisfy the following
criteria:

\begin{enumerate}
\item the secondary vertex appears as a kink without black prongs, nuclear
  recoils, blobs or Auger electrons;
\item the transverse momentum of the decay muon (ha\-dron) with respect to the
  parent direction is larger than 250 MeV/c (to eliminate decays of strange
  particles);
\item the kink, in the $0\mu$ channel, occurs within 3 plates downstream from
  the neutrino interaction vertex plate.  Because of the lower background, the
  kink sear\-ch in the muo\-nic decay channel was extended to 5 plates, with a
  gain in efficiency of about 8\%.
\end{enumerate}
    
No $\tau^-$ decay candidate has been found satisfying the selection criteria.
  
\section{Experimental check of the kink finding efficiency}

The kink finding efficiency, $\epsilon_{kink} $, has been eva\-luated by Monte
Carlo simulation and experimentally checked looking at hadron interactions and
dimuon events.

A sample of about 55~m of hadron tracks was scanned in emulsion.  In the decay
search procedure, 21 neutrino interaction events with a hadron interaction have
been detected. This result is in good agreement with the expected value of
($24\pm2$) from Monte Carlo simulation.  

Part of the dimuon sample (two muons on the final state) collected in 1995 and
1996 have been analysed searching for the reaction $\nu_\mu N\rightarrow\mu^-
D^+ X$ with the subsequent muonic decay of the $D^+$. Assuming a charm yield of
about $\sigma_{charm}/\sigma_{CC}\sim 5\%$, in the sample scanned we expect
$(22.8\pm 3.9)$ dimuon events and we found 25 events. This result shows that
Monte Carlo simulation and real data are in good agreement.

Although the number of events is too low to draw quantitative conclusions, we
can take these results as qualitative checks of the simulation of the automatic
scanning procedure.

\section { Sensitivity and backgrounds }

In this section we discuss the expected background from known sources and, in
absence of $\tau^-$ candidates, limits on the $\nu_{\mu}\rightarrow\nu_{\tau}$
oscillation parameters are derived. Both signal efficiencies and background
estimation have been evaluated from large samples of events, generated
according to the relevant processes, passed through a GEANT~\cite{geant} based
simulation of the detector response.  The output is then processed through the
same reconstruction chain as used for data. The simulated tracks in emulsion
are used to estimate the efficiencies of each scanning step. Apart from the
kink detection efficiency, only ratios of acceptances enter the calculation of
the experimental result and most of the systematic uncertainties of the
simulation cancel out.

\subsection { Background estimates }

Sources of background for the hadronic $\tau$ decay channel are:

\begin{itemize}
 
\item the production of negative charmed particles from the anti-neutrino
  components of the beam. These events constitute a background if the primary
  $\mu^+$ or $e^+$ remains unidentified. Taking into account the appropriate
  cross-sections and the bran\-ch\-ing ratios, in the present sample we expect
  $\sim0.02$ events from these sour\-ces;
  
\item the production of positive charmed mesons in char\-ged current
  interactions, if the primary lepton is not identified and the charge of the
  charmed particle daughter is incorrectly measured.  We expect $\sim 0.03$
  events from this source in the present sample;
  
\item the associated charm production both in charged (when the primary muon is
  lost) and neutral current interactions, when one of the charmed particles is
  not detected. The cross-sec\-tion for charm-anticharm pair production in
  neutral current interactions relative to the total charged current
  cross-section has been measured by the E531 experiment~\cite{chass} to be
  $0.13^{+0.31}_{-0.11}\%$.  The production rate of associated charm in
  char\-ged current interactions is unknown, but an upper limit is available
  ($<0.12\%$ ~\cite{chass}). In the present sample, the estimated background
  from this process represents $\lesssim 0.01$ events.
  
\item the main potential background to the hadronic $\tau^-$ decays is due to
  so-called hadronic ``white kin\-ks'', defined as 1-prong nuclear interactions
  with no heavily ionising tracks ({\em black} and {\em grey} tracks in
  emulsion terminology) and no evidence for nuclear break up (evaporation
  tracks, recoils, blobs or Au\-ger electrons).  Published data allowing to
  determine the white kink interaction cross-section are
  scarce~\cite{kek,balda}. In a dedicated experiment with 4 GeV pions at
  KEK~\cite{kek}, a very steep fall-off in $p_{T}$ was observed and only 1 out
  of 58 observed kinks had a $p_{T}$ larger than 300 MeV/c.
  
  Since the experimental information of the $p_{T}$ dependence is statistically
  poor at large values, a Monte Carlo simulation, based on a modified version
  of FLUKA~\cite{ferrari1, ferrari2}, has been performed. The results of this
  simulation are in good agreement with the $p_{T}$ dependence of the KEK
  measurement. An effective white kink mean free path in emulsion,
  $\lambda_{wk}\sim 22$ m, is obtained for a $p_{T}$ cut at 250 MeV/c, using a
  pion energy spectrum as observed in the 0-$\mu$ sample. The above result is
  compatible with the observation of 4 events with $p_{T}>$ 250 MeV/c, all at
  large distance from the primary vertex, along a total path of $\sim 92$~m of
  scan-back tracks, and corresponds to a background estimate of 0.5 events
  within 3 plates downstream from the primary vertex plate.  As the statistics
  will increase, this background source will be better measured using the
  CHORUS data. It will be possible to significantly reduce, by kinematical
  analysis, the effect of this background on the sensitivity to a possible
  oscillation signal.

\end{itemize}

\noindent 
The main source of potential background in the muonic $\tau$ channel is the
charm production.  We expect less than 0.1 events in the current sample from
the anti-neutrino components of the beam:

$$\bar{\nu}_\mu(\bar{\nu}_e)M\rightarrow \mu^+(e^+)D^-X$$

followed by

$$D^-\rightarrow\mu^-X^0$$

in which the $\mu^+(e^+)$ escapes the detection or is not identified.

The prompt $\nu_{\tau}$ contamination of the beam~\cite{prompt} is a background
  common to both the hadronic and muonic decay channels. For the
  present sample the expected background is much less than 0.1 events.

\subsection{ Oscillation sensitivity }

In the usual approximation of a two-flavour mixing scheme, the probability of
$\nu_{\tau}$ appearance in an initially pure $\nu_{\mu}$ beam can be expressed
as

\begin{eqnarray}
P_{\mu\tau}(E) & = & \sin2t \cdot \int \Psi(E,L)\cdot \nonumber \\
 & & \cdot sin^{2} \left(\frac{1.27 \cdot \Delta m^2_{\mu\tau}(eV^2)\cdot
          L(km)}{E(GeV)}\right) \cdot dL \nonumber
\end{eqnarray}

\noindent

where

\begin{itemize}

\item[-] $E$ is the incident neutrino energy;

\item[-] $L$ is the neutrino flight length to the detector;
  
\item[-] $\theta_{\mu\tau}$ is the effective $\nu_{\mu}-\nu_{\tau}$ mixing
  angle;
  
\item[-] $\Delta m^2_{\mu\tau}$ is the difference of the squared masses of the
  two assumed mass eigenstates;
  
\item[-] $ \Psi(E,L)$ is the fraction of $\nu_\mu$ with energy $E$
originating at a distance between $L$ and $L+dL$ from the emulsion target.
\end{itemize}

The $\tau^-$ channels considered in the $\nu_\mu\rightarrow\nu_\tau$
oscillation search we describe in this paper are: 

1) $\tau\rightarrow\mu$, 2)
$\tau\rightarrow h$, 3) $\tau\rightarrow e$ and 4) $\tau\rightarrow\bar{\mu}$
(the $\mu$ is not identified) channels.
 
The expected number, $N_{\tau i}$ ($i=1,2,3,4$), of observed $\tau^-$ decays into a channel of
bran\-ching ratio $BR_i$ is then given by

\begin{equation}
N_{\tau i} = BR_i \cdot \int \Phi_{\nu_{\mu}} \cdot P_{\mu\tau} \cdot \sigma_{\tau} \cdot
            A_{\tau i} \cdot \epsilon_{\tau i} \cdot dE
\end{equation}

\noindent

with

\begin{itemize}
\item[-] $BR_{(1~or~4)} = BR(\tau\rightarrow\nu_\tau\bar{\nu}_\mu \mu^-)  =
  (17.35\pm0.10)\%$~\cite{pdg}). 
\item[-] $BR_2 = BR(\tau\rightarrow\nu_\tau h^- nh^0) = (49.78\pm0.17)\%$~\cite{pdg});
\item[-] $BR_3 = BR(\tau\rightarrow\nu_\tau\bar{\nu}_e e^-) =
  (17.83\pm0.08)\%$~\cite{pdg}); 
\item[-] $\Phi_{\nu_{\mu}}$ the incident $\nu_{\mu}$ flux spectrum;
  
\item[-] $\sigma_{\tau}$ the charged current $\nu_{\tau}$ interaction cross-section;
  
\item[-] $A_{\tau i}$ the acceptance and reconstruction efficiency for the
  considered channel (up to the vertex plate location);

\item[-] $\epsilon_{\tau i}$ the corresponding efficiency of the decay search
  procedure;

\end{itemize}

With proper averaging (denoted by $\langle \rangle $), $N_{\tau i}$ can also be written as a
function of $n_i$:

\begin{equation}
N_{\tau i} = BR_i \cdot n_i \cdot  \langle P_{\mu\tau} \rangle \cdot \frac{ \langle\sigma_{\tau} \rangle}{\langle\sigma_{\mu} \rangle} \cdot
            \frac{ \langle A_{\tau i} \rangle}{\langle A_{\mu} \rangle}\cdot
            \langle\epsilon_{\tau i} \rangle
\end{equation}

\noindent    
where 

\begin{itemize}
\item[-] $n_1=N_\mu$ (the number of located charged current $\nu_{\mu}$
  interactions corresponding to the considered event sample) and $n_2=n_3=n_4=\left({N_{\mu}}\right)_{0-\mu}$ (the product of $N_{\mu}$ and the relative fraction of
the 0-$\mu$ sample for which the analysis has been completed);

\item[-] $\langle \sigma_{\mu(\tau)}\rangle = \int
  \frac{d\sigma_{\mu(\tau)}}{dE}\cdot\Phi_{\nu_{\mu}}\cdot dE$. It takes
  into account quasi-elastic interactions, resonance production and deep
  inelastic interactions ($\sigma(\frac{\langle\sigma_\tau\rangle}{\langle\sigma_\mu\rangle})_{syst}\sim7\%$);
\item[-] $\langle A_{\mu(\tau i)}\rangle = \int
  \frac{d\sigma_{\mu(\tau)}}{dE}\cdot
  A_{\mu(\tau i)}\cdot\Phi_{\nu_{\mu}}\cdot dE$\\ ($\sigma(\frac{ \langle
    A_{\tau i}
    \rangle}{\langle A_{\mu} \rangle})_{syst}\sim7\%$);
\item[-] $\langle\epsilon_{\tau i}\rangle$ is the average efficiency of the
  decay search procedure for the accepted events ($\sigma(\langle\epsilon_{\tau
    i}
  \rangle)_{syst}\sim10\%$);
\end{itemize}

To allow an easy combination of the results from the 1-$\mu$ and 0-$\mu$ event
samples, it is useful to define the ``equivalent number of muonic events'' of
the 0-$\mu$ sample by
 
\begin{equation}
N_{\mu}^{eq} =  \left({ N_{\mu}}\right)_{0-\mu} \cdot \sum_{i= 2}^{4}
               \frac { \langle A_{\tau i}\rangle} { \langle A_{\tau\mu}\rangle}
               \cdot \frac{\langle \epsilon_{\tau i} \rangle}{\langle \epsilon_{\tau\mu} \rangle}\cdot\frac{BR_i}{BR_\mu}
\end{equation}

The 90\% C.L. upper limit on the oscillation probability then simplifies to
 
\begin{equation}
P_{\mu\tau} \leq \frac {  2.38 \cdot r_{\sigma} \cdot r_{A} }
         {  BR_\mu\cdot \langle\epsilon_{\tau\mu}\rangle\cdot\left[ N_{\mu} +
          N_{\mu}^{eq} \right] }
\end{equation}

where $r_{\sigma} = \langle \sigma_\mu\rangle/\langle\sigma_\tau\rangle$ and
$r_{A} = \langle A_\mu\rangle/\langle A_{\tau\mu}\rangle$.  

In the above formula, the numerical factor 2.38 takes into account the total
systematic error (17\%) following the prescription given in~\cite{stat}.  The
systematic error is mainly due to the reliability of the Monte Carlo simulation
of the scanning procedures.

The estimated values of the quantities appearing in this expression are given
in Table~\ref{tab5}. No statistical errors are quoted since they are much
smaller than the systematic uncertainty.

Using the present sample the following 90\% C.L. limit is obtained

\begin{equation}
P_{\mu\tau} \leq 6.0\cdot 10^{-4}
\end{equation}      

\begin{figure}[tb]
  \begin{center}
    \resizebox{0.5\textwidth}{!}{
      \includegraphics{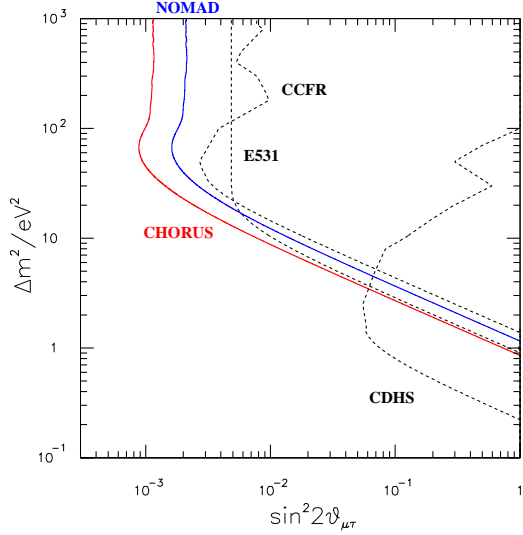}}
      \caption{Present result compared with the recent NOMAD
      result~\cite{nomad} (full line) and the previous limits (dotted lines).
        \label{fig:excl}}
  \end{center}
\end{figure}

In a two flavour mixing scheme, the 90\% C.L. excluded region in the ($\sin2t$,
$\Delta m^2_{\mu\tau}$) parameter space is shown in Figure~\ref{fig:excl}.
Maximum mixing between $\nu_\mu$ and $\nu_\tau$ is excluded at 90\% C.L. for
$\Delta~m_{\mu\tau}^2~\gtrsim~0.9~\mbox{eV}^2$; the large $\Delta m^2$ are
excluded at 90\% C.L. for $\sin2t>1.2\cdot10^{-3}$.

\begin{table*}
\begin{center}
\caption{\em {Quantities used in the estimation of the sensitivity} }
\label{tab5}
\begin{tabular}{llll}
\\
\hline
\hline
  & 1994 & 1995 & 1996 \\ 
\hline
$N_{\mu}$ &18,286 & 20,642 & 30,128\\
$r_{\sigma}$ & 1.89 & 1.89 & 1.89 \\
$r_{A}$ & 0.93 & 0.93 & 0.93 \\
${\langle A_{\tau\mu}\rangle}$ & 0.39 & 0.39 & 0.39\\
${\langle A_{\tau h}\rangle}$ & 0.17 & 0.17 & -\\
${\langle A_{\tau e}\rangle}$ & 0.093& 0.093 & -\\
${\langle A_{\tau\bar{\mu}}\rangle}$ & 0.026& 0.026 & - \\
$\langle\epsilon_{\tau\mu}\rangle$ & 0.53 & 0.35 & 0.37\\
$\langle\epsilon_{\tau h}\rangle$ & 0.24 & 0.25 & -\\
$\langle\epsilon_{\tau e}\rangle$ & 0.12 & 0.13 & -\\
$\langle\epsilon_{\tau\bar{\mu}}\rangle$ & 0.22& 0.23 & -\\
$N_{\mu}^{eq}$ & 11,987& 12,743 & - \\
 \hline
  \end{tabular}
\end{center}
\end{table*} 

\section {Conclusions }
  
The emulsion scanning methods, previously described in~\cite{1mupap,0mupap},
have been applied to a fraction of the 1994-1995-1996 data.  No $\tau^-$ decay
candidate has been found, leading to a more stringent 90\% C.L. upper limit on
the $\numtonut$ oscillation probability ($P_{\mu\tau} \leq 6.0\cdot10^{-4}$).
This negative result is compatible with an estimated background of less than
one event, mainly from ``white kink'' secondary interactions in the $0\mu$
channel.  With the large increase in statistics expected from the ongoing
analysis of the data, a direct measurement of this background process will be
possible and kinematical cuts are planned for its reduction.  Furthermore, a
second phase of the analysis (with better efficiencies, larger statistics and
faster automatic emulsion scanning) has started with the aim of reaching the
design sensitivity $(P_{\mu\tau} \leq 1.0\cdot10^{-4})$~\cite{proposal}.

\end{document}